\documentclass[12pt]{article}
\usepackage[letterpaper,margin=1in]{geometry}
\usepackage{setspace} 
\usepackage[utf8]{inputenc}
\usepackage{authblk,amsmath,amssymb,graphicx,tensor,braket,hyperref}
\usepackage[numbers,sort&compress]{natbib}
\usepackage[english]{babel}

\title{Bell's nonlocality and gravity}

\author{Yuri Bonder$^*$ and Johas D. Morales}
\affil{Instituto de Ciencias Nucleares\\
Universidad Nacional Aut\'onoma de M\'exico\\
Apartado Postal 70-543, Coyoac\'an 04510 Cd.Mx., M\'exico\\ $^*$Corresponding author, email: \href{mailto:bonder@nucleares.unam.mx}{bonder@nucleares.unam.mx}}
\date{Essay written for the Gravity Research Foundation\\
2022 Awards for Essays on Gravitation\\\vfill
}

\begin{document}

\maketitle

\begin{abstract}
The experimental results that test Bell's inequality have found strong evidence suggesting that there are nonlocal aspects in nature. Evidently, these nonlocal effects, which concern spacelike separated regions, create an enormous tension between general relativity and quantum mechanics. In addition, by avoiding the coincidence limit, semiclassical gravity can also accommodate nonlocal aspects. Motivated by these results, we study if it is possible to construct geometrical theories of gravitation that are nonlocal in the sense of Bell. We propose three constructions of such theories, which could constitute an important step towards our understanding of the interplay between quantum mechanics and gravitation.
\end{abstract}

\pagebreak

\section{General motivation}

It is an empirical fact that Bell's inequalities are violated. This, in turn, strongly suggests that there are nonlocal features in nature. What is more, this nonlocality relates spacelike separated regions of spacetime. At first sight, this reveals an almost insurmountable tension with general relativity (GR), where gravity is described by local geometrical entities that have the causal information. (For an excellent discussion on this issue see Ref.~\cite{maudlinnonlocality}, see also Ref.~\cite{Tumulka}). The logical conclusion seems to be that, without leaving the local/geometrical paradigm, we will never be able to construct a theory of gravitation that can account for matter in every possible state.

In this essay we show that it is in principle possible to build nonlocal theories of gravity that are still geometrical. To motivate our proposals, we first present a brief overview of Bell seminal work. We then argue that, within our most fundamental description of nature, namely, in semiclassical gravity, it is also possible to incorporate nonlocal effects. Subsequently, we turn to the question of how to describe gravity geometrically while, at the same time, it can accommodate nonlocal effects. It should be mentioned that the nonlocalities considered here are completely different from those found in the so-called ``nonlocal theories of gravitation'' where the nonlocalities are related to the formal appearance of infinitely many derivatives in the action, or they relate causally connected regions of spacetime \cite{nonlocalgravity}.

Clearly, addressing the question of nonlocality in gravity could constitute a crucial step towards a successful quantization of gravity. Nevertheless, the fact that we are trying to incorporate a nonlocal structure into the geometrical framework does not imply that we believe that gravity is fundamentally geometrical; we are simply tackling some questions that, we believe, could shed light into the quantum nature of gravity.

\section{Bell's theorem}

Entanglement is one of the most innovative and counterintuitive notions that arise in quantum mechanics. In fact, since the early years of quantum mechanics, the nature of entanglement was disputed; the best known example is the work by Einstein, Podolsky, and Rosen \cite{EPR}. The issues raised in these type of studies motivated Bell to study these issues, who, in 1964, presented his first theorem \cite{belloriginal}. This theorem states that, under fairly reasonable assumptions, any deterministic theory capable of reproducing the predictions of quantum mechanics must exhibit some kind of nonlocality. This paper was the first of several works where Bell refined his hypothesis, until he was able to include probabilistic theories, such as quantum mechanics \cite{Bell2004-BELSAU}. Bell's works can be summed up in an inequality that constrains the predictions of local theories. As it is well known, quantum mechanics, having the aforementioned entangled states, is nonlocal, and it violates Bell's inequalities.
 
It is worth mentioning that the assumptions of Bell's theoriem are quite general. Besides giving a precise mathematical notion of locality, Bell assumed ``statistical independence.'' This hypothesis assumes that the measurements are independent of the experimental composition that generates the entangled pair \cite{hipotesisbell}, which is essentially what is done every time the scientific method is applied. Moreover, it is important to emphasize that Bell's theorem can be applied in a wide range of circumstances \cite{maudlinnonlocality}, even outside the realm of quantum mechanics.

Experimental work to test Bell's inequality began in the early 1980s, using polarized photons \cite{Aspect1,Aspect2,Aspect3}. These experiments found that Bell's inequalities are indeed violated. What is more, over time there have been other experiments that have tested the inequality in extreme situations and that have eliminated gaps in the theorem's hypotheses \cite{discipulosdeaspect,bellexperimento,bellexperimento2,bellexperimento3,otroexperimento,experimentoelectrones}. The results, in all cases, is that the inequalities are violated, and hence, they provide evidence that the world has nonlocal features.

\section{Semiclassical gravity}

In quantum field theory in curved spacetimes \cite{waldrojo} people came up with a concrete recipe to build the energy-momentum tensor, for a given quantum field, such that, its renormalized expectation value can take the place of its classical counterpart in Einstein's equation. This scheme is known as semiclassical gravity and it is the best framework available to study the effects that quantum systems have on a classical spacetime. For the following part of the discussion we can focus on a real free Klein-Gordon field $\phi$ of mass $m$. In this case the energy-momentum tensor is given by\footnote{Throughout the essay, the notations a conventions of Ref.~\cite{waldgr} are used.}
\begin{equation}\label{KGtab}
T_{ab}=\nabla_a\phi\ \nabla_b\phi-\frac{1}{2}g_{ab}\left(\nabla_c\phi\ \nabla^c\phi+m^2\phi^2\right),
\end{equation}
where $g_{ab}$ and $\nabla_a$ are the spacetime metric and the associated derivative, respectively, and raised indexes indicate the contraction with an inverse metric, $g^{ab}$.

Importantly, in quantum field theory in curved spacetimes, $\phi$ only makes sense as an operator-valued distribution, and quadratic distributions, like $\phi^2$, are ill defined [cf. Eq.~\eqref{KGtab}]. Therefore, to have a well-defined energy-momentum tensor, one must appeal to a point splitting procedure, where the fields are evaluated at different points, and, in the end, one takes the limit where the two points coincide; this last step is known as the coincidence limit. However, the coincidence limit produces divergences that must be systematically removed. This process of removing divergences is called Hadamard renormalization and it has the undesirable feature that it selects a subset of states from the Fock space. In fact, to compute the renormalized expectation value of the energy-momentum tensor, only Hadamard states, which have the same divergent structure in the two-point function than the corresponding function in flat spacetime, can be utilized.

As we have already discussed, we are taking the point of view that nature is nonlocal. On the other hand, Hadamard's renormalization process, and the corresponding restriction in the Fock space, only play a role in if the coincidence limit is taken. Therefore, it is relevant to study if this limiting procedure can be avoided. In this scenario, the energy-momentum expectation value will depend on two spacetime points; this nonlocal object will be denoted by $\braket{T_{ab}}_\psi$. Note that $\braket{T_{ab}}_\psi$ depends on $g_{ab}$ and it is intimately related with the two-point functions \cite{fulling}. Therefore it is reasonable to expected that $\braket{T_{ab}}_\psi$ is sensitive to the causal spacetime structure. In what follows we assume that $\braket{T_{ab}}_\psi$ can be built out of the causal spacetime structure, and we focus on the possible nonlocal objects that could take the place of the Einstein tensor in a nonlocal Einstein-like equation.

\section{Nonlocal geometries}

The goal is to discuss what kind of objects can we use in the geometric side of an Einstein-like nonlocal equation. We require that all candidates to take this place have the appropriate limiting behavior. Namely, they must become the conventional Einstein tensor, $G_{ab}(x)$, in the coincidence limit to be able to make contact with semiclassical gravity. In addition, we need to have an object describing the causal spacetime structure. This last assumption is related to the properties of $\braket{T_{ab}}_\psi$, but it is also compatible with our motivation that the nonlocal effects are Bell-like, and hence, they concern spatially related spacetime regions.

Bitensors are geometrical objects that have a nonlocal character \cite{Poisson}. Concretely, a bitensor is a multilinear map that takes $k$ covectors and $\ell$ vectors in $x$, and $k'$ covectors and $\ell'$ vectors in $x'$, and produces a real number. Such an object can be represented in an abstract index notation by
$\tensor{T}{^{a_1}^{...}^{a_k}_{b_1}_{...}_{b_\ell}^{a'_1}^{...}^{ a'_{k'}}_{b'_1}_{...}_{b'_{\ell'}}}$, where the (un)primed indices act on vectors and covectors in ($x$)$x'$.

Perhaps the best known bitensor, which is well defined within a normal convex hull, is the parallel propagator $\tensor{g}{^{a'}_b}=\tensor{g}{^{a'}_b}(x,x')$. This object takes a vector in $x$ and produces the parallel transported vector at $x'$ along the geodesic joining $x$ and $x'$. To transport arbitrary tensors from $x'$ to $x$, the inverse of $\tensor{g}{^{a'}_b}$ is also needed, which is denoted by $\tensor{g}{^a_{b'}}=\tensor{g}{^a_{b'}}(x,x')$. Obviously $\tensor{g}{^a_{b'}}\tensor{g}{^{b'}_c}=\delta^a_c$; however, in general $\tensor{g}{^{a''}_{b'}}\tensor{g}{^{b'}_c}\neq\tensor{g}{^{a''}_c}$, where conventional contraction is defined for each type of index. It should also be mentioned that bitensors can be derived with respect to each of its two points; these derivatives are defined as standard (covariant) derivatives in the point of the derivation, and by keeping everything at other point fixed. Thus, derivatives associated to different points commute. Notice that, for any bitensor, one can take a ``coincidence'' limit $x'\to x$. Throughout this essay, we consider that such limits exist regardless of the direction in which $x'$ approaches $x$. 

The bitensors that replace $G_{ab}(x)$ in a nonlocal Einstein-like equation are called Einstein bitensors; we require that they coincide with $G_{ab}(x)$ in the coincidence limit. We turn to describe three concrete constructions of Einstein bitensors.

\subsection*{Conformal factor}

It is well known that, for a given metric $g_{ab} $ and a real smooth positive function $\bar{\Omega}$, the conformally related metric, $\bar{\Omega}^2 g_{ab}$, has the same causal structure than $g_{ab}$. The first of the strategies we put forward to construct an Einstein's bitensor is motivated by this simple observation. The idea is to allow the conformal factor to be a bifunction that will be determined when solving the corresponding equations of motion. The only requirement on $\Omega(x,x')$, besides being positive, is that $\Omega(x,x')\to1$ as $x'\to x$, fast enough to recover semiclassical gravity.

We can construct an Einstein bitensor by using the expressions for the curvature of a conformally related metric \cite{waldgr}. The outcome is
\begin{eqnarray}
G_{ab}(x,x')&=& G_{ab}(x) -2\nabla_a\nabla_{b}\ln\Omega(x,x') +2\nabla_{a}\ln\Omega \ \nabla_{b}\ln\Omega(x,x')\nonumber\\
&&+2g_{ab}\nabla_c\nabla^c\ln\Omega(x,x')+g_{ab}\nabla_c\ln\Omega\ \nabla^c\ln\Omega(x,x').\label{Einsteinconforme}
\end{eqnarray}
This $G_{a{b}}(x,x')$ satisfies all the requirements to be considered an Einstein's bitensor since, clearly, $\displaystyle{\lim_{x' \to x}}G_{a{b}}(x,x')=G_{ab}(x)$ and the causal structure is codified in $g_{ab}$. Thus, Eq.~\eqref{Einsteinconforme} is a concrete proposal that satisfies all the required properties.

\subsection*{Synge bifunction}

Another well-known bitensor is the Synge bifunction $\sigma(x,x')$ \cite{syngeGR}. This bifunction corresponds to the geodesic distance between $x$ and $x'$ (again, we work in a normal convex hull). Importantly, starting with $\sigma(x,x')$, we can produce new bitensors by taking derivatives. It turns out that $\displaystyle{\lim_{x' \to x}} \nabla_a \nabla_b \sigma(x,x')=g_{ab}(x)$; and there are similar relations obtained by taking derivatives at $x'$. Notably, this last relation has motivated some approaches where the metric is considered to emerge from $\sigma(x,x')$ \cite{achim}.

The proposal then is to build an Einstein bitensor using $ \nabla_a \nabla_b \sigma(x,x')$ in place of $g_{ab}$, without taking the coincidence limit. Clearly, by construction, this Einstein bitensor will reduce to the conventional Einstein tensor in such a limit. Moreover, the causal structure of spacetime is encoded in the sign of $\sigma(x,x')$. The main issue within this strategy is that one need to take the coincidence limit to some $ \nabla_a \nabla_b \sigma(x,x')$ to have a conventional metric (and its inverse), which is needed in $\braket{T_{ab}}_\psi$ and to make the contractions necessary to compute the corresponding Einstein bitensor. Nevertheless, it is clear that an Einstein bitensor with the desired properties can also be constructed form the Synge function.

\subsection*{Bitensorial connexion}

It is well known that GR accepts a variational formulation where the metric and the connection are \textit{a priori} independent \cite{Hehl,magdinamica}. Assuming further that the action can be separated into a gravity term, that is independent of the matter fields, and a matter part, $S_M$, then, the variation with respect to the metric, the nonmetric part of the connection, $\tensor{C}{^c_a_b}$, and the matter fields, $\varphi$, produces the equations of motion:
\begin{eqnarray}
G_{ab}(g,C)&=&8\pi T_{ab}(g,C,\varphi),\label{metricEOM}\\
g^{de}\tensor{C}{^{a}_{de}}\delta^{b}_c +\tensor{C}{^d_{dc}}g^{ab}-2\tensor{C}{^{a}_{ce}}g^{be}&=&16\pi\tensor{\Sigma}{_c^{ab}}(g,C,\varphi),\label{connectionEOM}\\
\frac{\delta S_M}{\delta\varphi}&=&0,
\end{eqnarray}
where $T_{ab}$ and $\tensor{\Sigma}{_c^a^b}$ are, respectively, the variations of $S_M$ with respect to $g_{ab}$ and $\tensor{C}{^c_a_b}$. Evidently, if $\tensor{\Sigma}{_c^a^b}=0$, then $\tensor{C}{^c_a_b}=0$, and GR is recovered. 

Unfortunately, to our knowledge, the semiclassical theory associated with Eqs.~\eqref{metricEOM}-\eqref{connectionEOM} has not been studied. Still, this semiclassical theory would require that $T_{ab}$ and $\tensor{\Sigma}{_c^a^b}$ are replaced by their expectation values. Thus, it is expected that both, $T_{ab}$ and $\tensor{\Sigma}{_c^a^b}$, must become nonlocal objects\footnote{If $S_M$ is the standard model action, only Dirac spinors produce a nontrivial $\tensor{\Sigma}{_c^a^b}$. Importantly, this $\tensor{\Sigma}{_c^a^b}$ is quadratic in the spinors, and thus, its expectation value is nonlocal in the same sense as $\braket{T_{ab}}_\psi$.}. Moreover, it is clear from the fact that Eq.~\eqref{connectionEOM} is simply an algebraic equation that, if we do not take the coincidence limit and we impose that the metric is a conventional tensor, then $\tensor{C}{^a_b_c}$ has to be a bitensor\footnote{In case some indexes in $\tensor{C}{^{a}_b_c}(x,x')$ are actually primed indexes, as we naively expect from simple geometrical considerations, one can use the parallel propagator bitensor to ``bring'' all indexes to $x$.} $\tensor{C}{^{a}_b_c}(x,x')$. The important point is that this bitensorial connection leads to an Einstein bitensor of the form $G_{ab}(x,x')=G_{ab}[g(x), C(x,x')]$. What is more, the semiclassical theory is automatically recovered in the coincidence limit since Eq.\eqref{connectionEOM} requires $\tensor{C}{^a_b_c}$ to be a conventional tensor.

Recall that the Riemann tensor associated with an arbitrary connection $\tensor{C}{^c_{ab}}$ satisfies
\begin{equation}\label{Riemann}
\tensor{R}{_a_b_c^d}=\tensor{R}{_a_b_c^d}(x)-2\nabla_{[a}\tensor{C}{^d_{b]c }}+2\tensor{C}{^e_{[a|c|}}\tensor{C}{^d_{b]e}},
\end{equation}
where $\tensor{R}{_a_b_c^d}(x)$ is the Riemann tensor associated with $g_{ab}$. Clearly, this gives rise to a Riemann bitensor when $\tensor{C}{^{a}_b_c}=\tensor{C}{^{a}_b_c}(x,x')$. One can then use the metric to build the corresponding Einstein bitensor, which has the form
\begin{eqnarray}
G_{ab}(x,x')&=&G_{ab}(x)  -2\nabla_{[a}\tensor{C}{^{c}_{c]b}} (x,x')+g_{ab}(x)g^{cd}(x)\nabla_{[c}\tensor{C}{^{e}_ {e]d}}(x,x')\nonumber\\
&&+2 \tensor{C}{^{d}_{[a|b|}}(x,x')\tensor {C}{^{c}_{c]}_d}(x,x') -g_{ab}(x)g^{cd}(x)\tensor{C}{^{e}_{[c|d| }}(x,x')\tensor{C}{^f_{f]e}}(x,x').
\end{eqnarray}
Again, this construction has the required properties of producing conventional semiclassical gravity in the coincidence limit and having the information on the causal spacetime structure. Moreover, by construction, this tensor satisfies Eq.~\eqref{metricEOM}, which is the Einstein-like equation, but where $T_{ab}$ must be replaced by its expectation value.

\section{Discussion}

In this essay, three methods are presented to construct an Einstein bitensor that can be used in nonlocal semiclassical gravity, \textit{i.e.}, in semiclassical gravity before taking the coincidence limit. The guide is to keep an object that describes the causal structure and to get the correct result in the coincidence limit. Here we presented concrete examples on how to use several nonlocal objects to build Einstein bitensors. However, all the constructions we presented have several ambiguities and there is no additional criteria to fix them. In trying to make the smallest possible departure from conventional semiclassical gravity, one can require the Einstein bitensors to be symmetric under the interchange of its indexes. This is strongly related to having a scheme that is symmetric when interchanging the two points in the bitensors. Constructing a point-symmetric bitensor is straightforward: one can always take one bitensor plus the bitensor with the points in the oposite order. This simple proposal could restrict the number of models, which, in turn, given the amount of decisions when building models, could be extremely valuable. Another requirement could be to have some sort of Bianchi identity, which does not seem to be present in the context of bitensors.

The mere fact that for a given point $x$ there is a ``random'' point $x'$ associated to it also seems puzzling. Perhaps this should be remedied by integrating in some spacetime region, say, the spacelike related region to $x$. Then, to have well-defined integrals, one would need to produce a conventional volume form. This seems to be feasible, but it seems unnatural in the models where the the natural volume form is a bitensor. This is simply because one would need to take the coincidence limit only for some of the bitensors. However, this is precisely the same issue that appears in the construction of the Einstein bitensor out of the Synge bifunction.

The immediate goal of this research plan is to construct a model that is mathematically self consistent to start making calculations. It is reasonable to expect that the calculations will be nontrivial, particularly as our bitensor toolbox is limited as compared to that of conventional tensors. Yet, two natural simplifications can be implemented. First, one can try to use a perturbation scheme in which, for example, one expands in powers of $\sigma(x,x')$. In this way, one would get the conventional local theory to zero order, and then, it seems possible to look for first order deviations. The second simplification is to study highly symmetric situations. In fact, we could start by working in maximally symmetric spacetimes where there are methods to compute $n$-point functions using bitensors \cite{Jacobson} (see also Refs. \cite{allenespinores,rourabitensores}).

The long-term objective is to select a handful of promising models to be analized in depth. These studies should tackle mathematical and phenomenological questions. In the mathematical front, one should try to extend some of the well-known GR notions. For example, given a fair definition of the energy-conditions for $\braket{T_{ab}}_\psi$, one could study if the conclusions of the singularity theorems can be avoided. Or, given that, by construction, the causal spacetime structure is conventional and hence one can use definitions such as global hyperbolicity, one can verify if the dynamics of these nonlocal theories is well-posed as an initial value problem. Regarding the phenomenology, recall that these models reduce to GR whenever $\braket{T_{ab}}_\psi$ behaves as a conventional energy-momentum tensor. Thus, one should not expect new signals in the usual tests of GR (solar system tests, gravity waves, \textit{etc.}). Perhaps the place to look for phenomenological effects is in cosmology, where there is plenty of data to analyze, and some apparent tensions \cite{DIVALENTINO2021102605} to resolve.

\section{Conclussions}

The motivation of our proposal stems from the results of Bell's work and the subsequent experiments testing Bell's inequalities. In addition, semiclassical gravity, without taking the coincidence limit, produces a nonlocal theory, which is natural in the sense that it does not impose restrictions in the Fock space. Thus, we propose that a more accurate description of nature could be performed using nonlocal geometrical objects, \textit{e.g.}, bitensors. We present the three constructions and briefly discuss some of their potential issues.

The research program outlined here is at a very early stage. Yet, we are convinced that it could produce very valuable lessons. If solutions with some clear physical interpretation are found, we could gain insight into the question of how entangled states gravitate. In turn, this should shed invaluable light to construct a satisfactory theory of quantum gravity. Another possible conclusion is that there is simply no way to generate a self consistent nonlocal geometrical theory. This would strengthen the argument that gravity, at a more fundamental level, must be nongeometrical \cite{YuriChryssDaniel}. Whatever the results end up being, we are proposing a very exciting and well-motivated research program, with the potential to illuminate some of the most obscure ---yet relevant--- corners of modern physics.

\subsection*{Acknowledgment}
This research was financially supported by UNAM-DGAPA-PAPIIT Grant IG100120, CONACyT FORDECYT-PRONACES grant 140630, and the CONACyT scholarships program.


\begin{thebibliography}{29}
\providecommand{\natexlab}[1]{#1}
\providecommand{\url}[1]{\texttt{#1}}
\expandafter\ifx\csname urlstyle\endcsname\relax
  \providecommand{\doi}[1]{doi: #1}\else
  \providecommand{\doi}{doi: \begingroup \urlstyle{rm}\Url}\fi

\bibitem[Maudlin(2011)]{maudlinnonlocality}
T.~Maudlin.
\newblock \emph{Quantum Non-Locality and Relativity: Metaphysical Intimations
  of Modern Physics}.
\newblock Wiley, third edition, 2011.

\bibitem[Tumulka(2006)]{Tumulka}
R.~Tumulka.
\newblock Relativistic version of the {G}hirardi--{R}imini--{W}eber model.
\newblock \emph{J. Stat. Phys.}, 125:\penalty0 821, 2006.

\bibitem[Mashhoon(2017)]{nonlocalgravity}
B.~Mashhoon.
\newblock \emph{Nonlocal Gravity}.
\newblock Oxford University Press, 2017.

\bibitem[Einstein et~al.(1935)Einstein, Podolsky, and Rosen]{EPR}
A.~Einstein, B.~Podolsky, and N.~Rosen.
\newblock Can quantum-mechanical description of physical reality be considered
  complete?
\newblock \emph{Phys. Rev.}, 47:\penalty0 777, 1935.

\bibitem[Bell(1964)]{belloriginal}
J.S. Bell.
\newblock On the {E}instein {P}odolsky {R}osen paradox.
\newblock \emph{Physics Physique Fizika}, 1:\penalty0 195, 1964.

\bibitem[Bell(2004)]{Bell2004-BELSAU}
J.S. Bell.
\newblock \emph{Speakable and Unspeakable in Quantum Mechanics: Collected
  Papers on Quantum Philosophy}.
\newblock Cambridge University Press, 2004.

\bibitem[Clauser and Horne(1974)]{hipotesisbell}
J.F. Clauser and M.A. Horne.
\newblock Experimental consequences of objective local theories.
\newblock \emph{Phys. Rev. D}, 10:\penalty0 526, 1974.

\bibitem[Aspect et~al.(1981)Aspect, Grangier, and Roger]{Aspect1}
A.~Aspect, P.~Grangier, and G.~Roger.
\newblock Experimental tests of realistic local theories via {B}ell's theorem.
\newblock \emph{Phys. Rev. Lett.}, 47:\penalty0 460, 1981.

\bibitem[Aspect et~al.(1982{\natexlab{a}})Aspect, Dalibard, and Roger]{Aspect2}
A.~Aspect, J.~Dalibard, and G.~Roger.
\newblock Experimental test of {B}ell's inequalities using time-varying
  analyzers.
\newblock \emph{Phys. Rev. Lett.}, 49:\penalty0 1804, 1982{\natexlab{a}}.

\bibitem[Aspect et~al.(1982{\natexlab{b}})Aspect, Grangier, and Roger]{Aspect3}
A.~Aspect, P.~Grangier, and G.~Roger.
\newblock Experimental realization of {E}instein-{P}odolsky-{R}osen-{B}ohm
  gedankenexperiment: {A} new violation of {B}ell's inequalities.
\newblock \emph{Phys. Rev. Lett.}, 49:\penalty0 91, 1982{\natexlab{b}}.

\bibitem[Tittel et~al.(1998)Tittel, Brendel, Zbinden, and
  Gisin]{discipulosdeaspect}
W.~Tittel, J.~Brendel, H.~Zbinden, and N.~Gisin.
\newblock Violation of {Bell} inequalities by photons more than 10 km apart.
\newblock \emph{Phys. Rev. Lett.}, 81:\penalty0 3563, 1998.

\bibitem[Weihs et~al.(1998)Weihs, Jennewein, Simon, Weinfurter, and
  Zeilinger]{bellexperimento}
G.~Weihs, T.~Jennewein, C.~Simon, H.~Weinfurter, and A.~Zeilinger.
\newblock Violation of {B}ell's inequality under strict {E}instein locality
  conditions.
\newblock \emph{Phys. Rev. Lett.}, 81:\penalty0 5039, 1998.

\bibitem[Gallicchio et~al.(2014)Gallicchio, Friedman, and
  Kaiser]{bellexperimento2}
J.~Gallicchio, A.S. Friedman, and D.I. Kaiser.
\newblock Testing {B}ell's inequality with cosmic photons: {C}losing the
  setting-independence loophole.
\newblock \emph{Phys. Rev. Lett.}, 112, 2014.

\bibitem[Giustina et~al.(2015)]{bellexperimento3}
M.~Giustina et~al.
\newblock Significant-loophole-free test of {B}ell's theorem with entangled
  photons.
\newblock \emph{Phys. Rev. Lett.}, 115, 2015.

\bibitem[Shalm et~al.(2015)]{otroexperimento}
L.K. Shalm et~al.
\newblock Strong loophole-free test of local realism.
\newblock \emph{Phys. Rev. Lett.}, 115, 2015.

\bibitem[Hensen et~al.(2015)]{experimentoelectrones}
B.J. Hensen et~al.
\newblock Loophole-free {B}ell inequality violation using electron spins
  separated by 1.3 kilometres.
\newblock \emph{Nature}, 526:\penalty0 682, 2015.

\bibitem[Wald(1995)]{waldrojo}
R.M. Wald.
\newblock \emph{{Quantum Field Theory in Curved Space-Time and Black Hole
  Thermodynamics}}.
\newblock University of Chicago Press, 1995.

\bibitem[Wald(1984)]{waldgr}
R.M. Wald.
\newblock \emph{{General Relativity}}.
\newblock University of Chicago Press, 1984.

\bibitem[Fulling(1989)]{fulling}
S.A. Fulling.
\newblock \emph{Aspects of Quantum Field Theory in Curved Spacetime}.
\newblock Cambridge University Press, 1989.

\bibitem[Poisson(2004)]{Poisson}
E.~Poisson.
\newblock The motion of point particles in curved spacetime.
\newblock \emph{Liv. Rev. Rel.}, 7, 2004.

\bibitem[Synge(1960)]{syngeGR}
J.L. Synge.
\newblock \emph{Relativity: {T}he General Theory}.
\newblock North-Holland Publishing Company, 1960.

\bibitem[Kempf(2021)]{achim}
A.~Kempf.
\newblock Replacing the notion of spacetime distance by the notion of
  correlation.
\newblock \emph{Frontiers in Physics}, 9, 2021.

\bibitem[Hehl et~al.(1995)Hehl, Mccrea, Mielke, and Ne’eman]{Hehl}
F.W. Hehl, J.D. Mccrea, E.W. Mielke, and Y.~Ne’eman.
\newblock Metric affine gauge theory of gravity: {F}ield equations, {N}oether
  identities, world spinors, and breaking of dilation invariance.
\newblock \emph{Phys. Rep.}, 258:\penalty0 1, 1995.

\bibitem[Vitagliano et~al.(2011)Vitagliano, Sotiriou, and
  Liberati]{magdinamica}
V.~Vitagliano, T.P. Sotiriou, and S.~Liberati.
\newblock The dynamics of metric-affine gravity.
\newblock \emph{Ann. Phys.}, 326:\penalty0 1259, 2011.

\bibitem[Allen and Jacobson(1986)]{Jacobson}
B.~Allen and T.~Jacobson.
\newblock Vector two-point functions in maximally symmetric spaces.
\newblock \emph{Commun. Math. Phys.}, 103:\penalty0 669, 1986.

\bibitem[Allen and Lutken(1986)]{allenespinores}
B.~Allen and C.A. Lutken.
\newblock {Spinor Two Point Functions in Maximally Symmetric Spaces}.
\newblock \emph{Commun. Math. Phys.}, 106:\penalty0 201, 1986.

\bibitem[Perez-Nadal et~al.(2010)Perez-Nadal, Roura, and
  Verdaguer]{rourabitensores}
G.~Perez-Nadal, A.~Roura, and E.~Verdaguer.
\newblock Stress tensor fluctuations in de {S}itter spacetime.
\newblock \emph{JCAP}, 2010.

\bibitem[{Di Valentino} et~al.(2021)]{DIVALENTINO2021102605}
E.~{Di Valentino} et~al.
\newblock Snowmass2021 - {L}etter of interest cosmology intertwined {II}: {T}he
  {H}ubble constant tension.
\newblock \emph{Astropart. Physics}, 131:\penalty0 102605, 2021.

\bibitem[Bonder et~al.(2018)Bonder, Chryssomalakos, and
  Sudarsky]{YuriChryssDaniel}
Y.~Bonder, C.~Chryssomalakos, and D.~Sudarsky.
\newblock Extracting geometry from quantum spacetime.
\newblock \emph{Foundations of Physics}, 48:\penalty0 1038, 2018.

\end{thebibliography}
\end{document}